\documentclass{jkas}

\def\year{2019} 
\def\month{August} 
\def\volume{52} 
\def\issue{4} 
\def\beginpage{99} 
\def\endpage{108} 
\def\received{May 4, 2019} 
\def\accepted{June 11, 2019} 
\date{Received \received; accepted \accepted}

\usepackage{flushend} 

\usepackage[colorlinks,
            pdftex,
            breaklinks,
            linkcolor=blue,
            urlcolor=blue,
            anchorcolor=blue,
            menucolor=blue,
            citecolor=blue]{hyperref}

\title{
The Evolution of Astronomical Observatory Design
}

\author[1,2]{Miguel~\'Angel~Castro Tirado}
\author[1,2]{Alberto~J.~Castro-Tirado}

\affil[1]{Instituto de Astrof\'isica de Andaluc\'ia (IAA-CSIC), Glorieta de la Astronom\'ia s/n, 18008 Granada, Spain; \email{macastrotirado@gmail.com}}
\affil[2]{Unidad Asociada al CSIC Departamento de Ingenier\'ia de Sistemas y Autom\'atica, Escuela de Ingenier\'ia Industrial, Universidad de M\'alaga, Arquitecto Francisco Pe\~nalosa, 29071 M\'alaga, Spain; \email{ajct@iaa.es}}

\def\corrauthor{
M. A. Castro-Tirado
}

\def\runningauthor{
M. A. Castro Tirado \& A. J. Castro-Tirado
}

\def\runningtitle{
The Evolution of Astronomical Observatory Design
}

\def\keywords{
history and philosophy of astronomy --- telescopes
}

\def\abstracttext{
This work addresses the development of the astronomical observatory all through history, from an architectural point of view, as a building in relation to the observing instruments and their 
functioning as a heterogeneous work center. We focused on 32 observatories (in the period 1259--2007) and carefully analyzed the architectures. Considering the impact of the 
construction itself or its facilities on the results of the research (thermal or structural stability, poor weather protection, turbulence, etc.), there is little attention paid to theories or studies of the architectural or construction aspects of the observatories. Therefore, this work aims to present a theoretical-critical contribution that, at least, invites the reflection of those involved in the development of astronomical observatories in the future.
}

\begin{document}
\jkashead 


\section{Introduction\label{sec:intro}}

The theoretical approach to astronomical observatories has its origin in the 
roots of the modern observatory \citep{brahe1602,caramuel1678}. However, 
historical-descriptive works that consider a particular project are the most 
common during the next centuries. In addition, some studies appeared focused 
on specific periods \citep{sayili1960,kwan2012,leverington2017}, others with 
a more general retrospective nature \citep{donnelly1971,krisciunas1999,
wolfschmidt2008} and only a few works focused on the 
architectural-constructive aspects of the observatories 
\citep{dumitrache2009,waumans2013}.

The concept of astronomical observatory itself lacks a precise and globally accepted definition given that historically there has not been a critical-intellectual movement that theorized about its nature or conception. Although it may seem obvious, what is understood by an observatory has changed over the years, and, in addition, it depends on the perspective from which it is valued. In this paper, the intention is to present an approach from the point of view of the architecture of the building itself, so that other aspects such as its astronomical equipment or its possible scientific usage will be set aside. That is to say, we will study the particularities related to the construction in which the instruments are located and whose purpose is the astronomical function. The analysis will also be limited to optical observatories.

Talking about evolution necessarily implies the extension of the study over time to analyze and assess how, when and why the changes that have defined the development of these astronomical centers have occurred. For this reason, following 
an earlier work \citep{castrotirado2019a}, starting from the examination of plans, ruins and the buildings themselves, this study goes back to the genesis of the observatories trying to illustrate the reciprocal relationship that exists between the design strategies of the building and the practice of its astronomical function, with the goal of theorizing about their future development. We have focused on 32 observatories which we consider to be representative ones in the period 1259--2007, deeply analyzing the architectural drafts.

\section{Origins\label{sec:origins}}

Astronomy is one of the oldest scientific disciplines of mankind. Since man raised his gaze to the sky and contemplated the stars with his own eyes, a relationship of fascination and devotion was established. This would be reflected in one way or another in different cultures that would arise across the planet. And even without having tools of precision or a basic understanding of the physical laws that rule the functioning of the universe, the first civilizations already had certain notions of the main astronomical facts. All based, always, on mere observations.

In any case, it is not possible to determine the date on which astronomy arises, although its roots probably extend to the origins of the human societies. What does seem clear is that, regardless of a spiritual or pseudo-religious sense, the beginnings of this science point towards an instrumentalization of the sky as a system for measuring periods of time and as a mean of orientation.
Beyond the numerous artistic representations of the Sun, the Moon and the stars
 that have been found \citep{pannekoek1961,pasztor2007}, sometimes more realistic and others more allegorical or figurative, or of the written testimonies that have lasted over time, this astronomical predilection remains through the incidence of some celestial phenomena in the disposition of certain archaeological sets.

Although certain remains of ancient civilizations show an impact of astronomy on their culture, it will be the proto-observatories that stand out as points of interest to know the way in which these 
peoples approached the practice of this science. In these spaces the astronomical observatories have their deepest antecedents and roots.
Although the astronomical principles that underlie the constitution of these assemblies are undoubted, there is no evidence to support the existence of research or scientific work in these places.

The oldest site that has been recorded so far is the Goseck Circle, in Germany, 
a solar proto-observatory with an annular plan that consists of a Neolithic 
structure dating back almost 7000 years \citep{brown2016,scherrer2018}. 
Of a similar period is the megalithic group of Nabta Playa, in Egypt. There 
are many other cases of lesser antiquity, among which the temple of Mnajdra, 
in Malta, or the famed Stonehenge, in England \citep{scherrer2018}. All these 
sites belong to ancient societies, with a low degree of technology and even, frequently, of nomadic habits. 
These facts propitiate relatively simple identifications of their constructions. In addition, the scale and materials of these structures favor their durability up to the present, allowing them to be studied and identified, where relevant, as proto-observatories.

However, the evolution towards more advanced civilizations tend to create settlements of multiple buildings that do not reach the precise degree of specialization for the observation function as such. 
Since no scientific instruments had been developed that required demanding conditions, systematic surveying of the sky could be carried out with the naked eye in almost any construction or, even, in the vicinity of these out in the open.

The oldest building strictly dedicated to astronomy of which there is evidence is Cheomseongdae, in South Korea. It was designed as a small bottle-shaped hollow tower made of masonry. 
Its function was to enable the use of astronomical instruments from an elevated position to avoid surrounding obstacles. 
It dates from the seventh century and remains in a great state of preservation 
\citep{park2010} and its square top and round body reflect the astronomical 
concept of the time (i.e. the Earth was considered a square and the sky was 
considered to have a round shape) with the 29 layers of stone from the bottom to the top 
corresponding to the 29.5 days of full lunation (or synodic month, i.e. the time 
from one new moon to the next) and the 27 layers of the round body 
representing the 27.3-day sidereal month, amongst other astronomical 
symbolism \citep{park2008}.

Anyway, although references to the Tower of Babel or to some type of 
astronomical institution directed by Ptolemy in Alexandria can be found 
\citep{mundt1927,george2005}, there is no evidence to support the existence as 
such of pre-medieval astronomical observatories.

\section{The Medieval Islamic Observatory\label{sec:Islamic}}

Coinciding with the beginning of the Middle Ages, a period of cultural 
darkness began in the Western world from which astronomy would not escape. 
During this period, Islamic culture came to play an essential role in the 
history of astronomy by preserving the vast Greek knowledge of this subject 
and incorporating its own findings. All this would be recovered for the West 
by the School of Translators of Toledo of Alfonso X, during the thirteenth 
century \citep{gargatagli1999}. 

The medieval Islamic observing sites can indeed be called observatories because of two main reasons: i) the fact that they were spaces dedicated specifically 
to the collective and prolonged study of celestial phenomena and 
ii) to the exchange of scientific knowledge.
It can be said that the observatory, as such, appears for the first time in 
this civilization. This is not a coincidence; the features 
of the Islamic culture as a society link with the observatory as a scientific 
institution \citep{sayili1960}. 
However, although this permanence is implemented in the Islamic 
observatory, they would have a reduced life \citep{sayili1960}.

Two singularities that would ultimately be decisive for the relevance of the 
Islamic observatory as an institution were the size of the observing tools and its 
patronage. The fact that the tools of observation and study reached a dimension 
such as to prevent their portability led to a necessary settlement. Royal or state support meant the formation of institutions 
larger and longer lived than an individual scientist \citep{sayili1980}. 

Although several observing posts, more or less temporary, had previously been 
established, it is difficult to speak of observatories as an institution 
before Al Mam\^{u}n (ninth century). The numerous observations collected by 
different authors allow us to recognize centers established in Shammasiya, 
Baghdad and Mount Qasiyun, Damascus \citep{mujani2012}. 
However, the fact that there are no remains, descriptions or illustrations of observatories could mean that no specific space was devoted to this use. This lack of material legacy makes it extremely difficult to distinguish between the observatories of this period and the specific observing posts established for an ephemeral study. 
The main differences amongst medieval islamic observatories were in the number of personnel dedicated to them, the quality of observing equipment and the life span of these centers.
That is to say, at this initial moment the observatories were not a clearly defined institution and it is complicated to indicate which observation sites transcended the barrier between the most advanced temporary work post and the embryonic scientific center. 
It would be from this stage onward that the Islamic observatory began to develop as a specialized astronomical institution linked to a scientific team and a specific location. It then begins to be structured with its own administration around a work program and its instruments become more precise, larger and heavier.

All these circumstances suppose a change in the paradigm of the Islamic observatory that, in a hesitant process over several centuries, transcends from a semi-improvised observation post to the scientific reference center of its time.
The construction of the observatory of Maragheh, in northwest Iran, was completed in 1264 and is part of the period that represents the climax of the evolution of the Islamic observatory \citep{saliba1987}.
In fact, the main progress and features of this institution in the Islamic world appear with this observatory, which became the paradigm to imitate of its time.

Although only the ruins of the construction remain, these together with the 
existing testimonies \citep{sayili1960} and the subsequent observatories built in its image, allow a description of the building to be developed. The observatory was a circular wall erected around a large central space occupied by the large quadrant for meridian transit observations. On both sides of this area there would be spaces to house the library, workshops, stores and other functions distinct from mere observation. Even though the date of the fall of the observatory at Maragheh is not exactly known, it still became the model to imitate by the great Islamic observatories that would arise during the following centuries, such as those of Samarkand and Istanbul.

\begin{figure*}[tp]
\centering
\includegraphics[width=160mm]{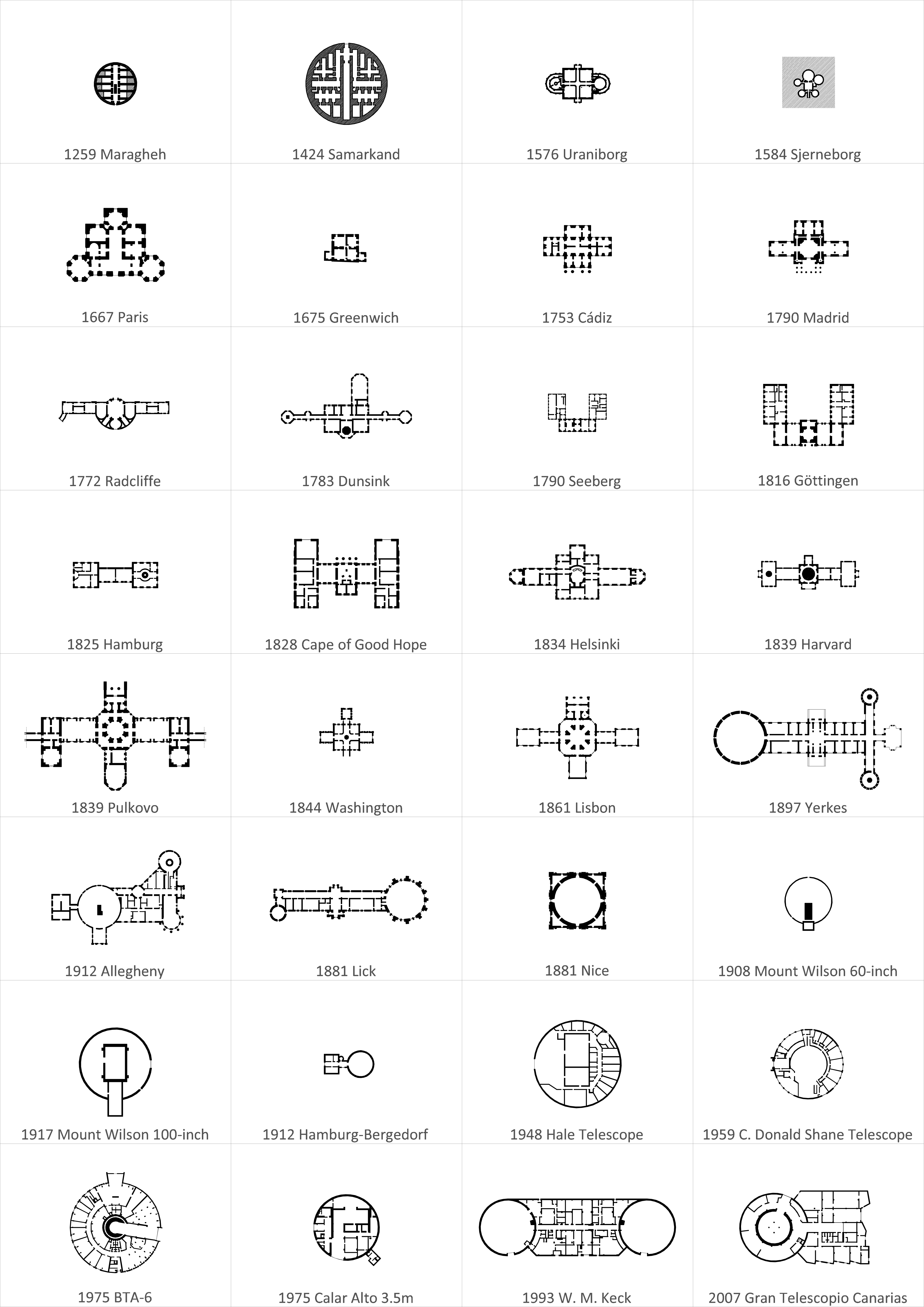}
\caption{Lower or access levels of the case studies considered for the analysis of the programmatic-functional configuration in its evolution from the Islamic-medieval observatory to the contemporary one. The scale is the same in order to allow direct comparison. (Drawings by M. A. Castro-Tirado)\label{fig:jkasfig1}}
\end{figure*}

\section{The Modern Observatory\label{sec:Modern}}

Although some references attribute the first European observatory to either 
Regiomontanus \citep{dreyer1890,gunther1932} or Copernicus \citep{zinner1943}, 
there is no evidence of a construction or of fixed instruments that support these 
claims \citep{zinner1943}. There is a consensus to 
place the first modern observatory in the palace of Wilhelm IV of Hesse-Kassel  
(this is recognized by \citealt{todd1922,pannekoek1961} or \citealt{pantin1999}). 
However, a platform built as a lookout point in the palace where portable 
instruments are taken out \citep{gaulke2009} does not represent an advance 
from the perspective of architecture. This does not invalidate its importance 
since it was a visit to these facilities that inspired Tycho Brahe, a few 
months later, to build the first two modern observatories 
\citep{zinner1956} in Hven characterized by having a certain degree of intention 
and architectural specialization for astronomy.

The first project of Tycho Brahe would be Uraniborg (1576), a palace in which the study of alchemy and astronomy would coexist with his own home. 
A dialogue between science and architecture in which the spatial configuration 
would adapt to the needs of research resulted in a solution that relegated 
alchemy to the basement and elevated astronomy to the upper floor, leaving the 
work and residence areas in the middle floor. 
Elevated platforms were built on slender poles from which it was possible to look at the sky without obstacles with covers that 
protected both the instruments and observers from poor weather conditions.

After several years Tycho Brahe noticed that the vibration of the platforms was being transmitted to the instruments and damaging the research unacceptably, so he planned a second center in which neither structural nor thermal instability would compromise astronomy. Following this idea Stjerneborg (1584) arose as a building excavated in the earth in which the observation tools would be fixed on pillars founded on the bedrock itself.  

The appearance of the telescope in the early seventeenth century and its early connection with astronomy would condition the future design of observatories, in which the increase in accuracy would require improvements to avoid architecture from becoming a brake on scientific research. However, since the first telescopes were small and manageable, many astronomers could make their observations from the windows, balconies, terraces or gardens of their own residence, or from public spaces. Possibly, this is why, during this period, the development of astronomical centers tends towards simple structures or, directly, to adding platforms to existing buildings from which to get better views of the sky, as in the universities of Leiden (1632) and Ingolstadt (1637) or in the Danish 
Rundet\aa rn (1637) \citep{donnelly1971}. Therefore, many years would pass between the appearance of the first modern observatory and the consolidation of the idea that a center for scientific observation required a building expressly designed for that purpose.

The observatory of Paris (1667) was created with the support of Louis XIV as a 
national institution in a context of increasing interest for culture and 
science. In addition to its scientific purpose, the project was to accommodate the headquarters of the Academy of Sciences and to enhance the prestige of its promoter. The resulting building is a palace where the institutional function shares space with the residences and the areas dedicated to astronomy: observation rooms, a zenith study room in the basement and a flat walkway for instruments.
For more than a century, observatories followed the example of the French, limited to  more or less contrived buildings that combined residence, work area and spaces where telescopes could operate. This is how the high observation rooms with vertical windows were reproduced, such as at the Greenwich Observatory (1675) or the openings from north to south were implemented at the Radcliffe Observatory (1772).

It was not until the construction of the Dunsink Observatory (1785) that 
significant changes were assimilated, where the architectural configuration 
was defined by the astronomical instruments that were to be installed. 
The person responsible for this project defined the key elements of the 
building as "situation, foundation and soil" \citep{ussher1787}. 
So the location was selected for its good visibility and accessibility, and the importance of stability meant the construction of foundations for the instruments to become structurally independent of 
the rest of the building. Systems were also developed to promote thermal stability in observation rooms. The design even reserves a privileged position to the telescopes, placing them under movable domes. 
The modern observatory is designed and planned to respond to a specific astronomical use during the period between the seventeenth and eighteenth centuries. The new observatories implement changes derived from the development of astronomy itself as a science (especially in relation to its instruments) and improve its performance for the sake of an optimal performance of the research activity without underestimating the welfare of its occupants.

Sometimes for strategic purposes (because of its importance for maritime navigation), sometimes for cultural ambition and in others, perhaps, by imitation, these buildings began to emerge throughout Europe in the eighteenth century, from where they would end up spreading to the rest of the world. Spaces for observation, characterized by the instruments that occupied them, coexisted with spaces in which the activities derived from these observations would take place (calculation, data analysis, discussion of results ...), secondary ones (preparation of instruments, offices, classrooms, libraries ...), and others corresponding to service uses (residences, storage areas ...) that, on occasion, moved to one or more independent buildings.
Likewise, the scientific and technological progress of this period had an impact on the observatories: 
on the one hand the buildings incorporated 
new functions such as physics or chemistry laboratories, photographic work 
rooms or optical and mechanical workshops, while on the other hand the work 
of astronomers was facilitated by mechanizing the movement of domes and 
telescopes. In addition, technical advances allowed a growth of the refractor 
that led to a race to having the largest telescope in the world. This struggle 
reached its peak at the Yerkes Observatory (1897) where 
the size of the refractor (1.02 m diameter) entailed an increase in the 
architectural scale of the building \citep{hussey1897}.

Although in this period the trend led to the observatories to move away from the city and, therefore, the town, some centers would still emerge, such as the Allegheny (1912) or the Griffith (1925), which would open to the public offering the use of some instruments and other activities to disseminate and promote astronomy.

\begin{table*}[t!]
\caption{Percentages of surface areas of observatory buildings dedicated to different functions.\label{tab:jkastable1}}
\centering
\begin{tabular}{lrrrr}
\toprule
Observatory        & Area (m$^2$)  & Observation (\%) & Derivatives (\%) & Accessories (\%) \\
\midrule
Maragheh (1259)     &  1164  &  4  &  96  &  0   \\
Samarkand (1424)  &  3082  &  3  &  97  &  0   \\
                   &        &     &      &      \\
Uraniborg (1576)   &  1180  &  9  &  22  & 69   \\
Stjerneborg (1584) &    91  & 71  &  29  &  0   \\
Paris (1667)       &  3152  & 48  &   8  & 44   \\
Greenwich (1675)   &   501  & 25  &   4  & 71   \\
C\'adiz (1753)     &  1364  & 33  &  39  & 28   \\
Madrid (1790)      &   618  & 27  &  14  & 59   \\
Radcliffe (1772)   &   643  & 36  &  32  & 32   \\
Dusink (1783)      &   802  & 21  &  27  & 52   \\
                   &        &     &      &      \\
Seeberg (1790)     &   473  & 13  &  4   & 83   \\
G\"ottingen (1816) &  1092  & 16  & 16   & 68   \\
Hamburgo (1825)    &   627  & 18  & 19   & 63   \\
Cape of Good Hope (1828) & 355 & 26 & 3  & 61   \\
Helsinki (1834)    &  1097  & 23  & 16   & 61   \\
Harvard (1839)     &   744  & 29  & 20   & 51   \\
Pulkovo (1839)     &  1429  & 44  & 18   & 38   \\
U.S. Naval in Washington (1844) & 367 & 26 & 40 & 34 \\
Lisboa (1861)      &   893  & 40  & 10   & 50   \\
Yerkes (1897)      &  2396  & 28  & 40   & 32   \\
Allegheny (1912)   &   969  & 44  & 26   & 30   \\
                   &        &     &      &      \\
Lick (great refractor, 1881)  &  957   &  43 & 32   & 25   \\
Niza (large equatorial, 1881) & 1040 & 43 & 43 & 14 \\
Mount Wilson (60-inch, 1908)  &  975 & 52 & 48 &  0 \\
Mount Wilson (100-inch, 1917) & 2509 & 51 & 49 &  0 \\
Hamburg-Bergedorf (large refractor, 1912) & 562 & 32 & 56 & 12 \\
Palomar (Hale telescope, 1948)& 1443 & 28 & 63 & 9 \\
Lick (Shane reflector, 1959)  & 2108 & 33 & 60 & 7 \\
SAO (BTA-6m, 1975) &   4480   & 33   & 60 & 7 \\
Calar Alto (3.5m, 1975) & 3709 & 18 & 74 & 8 \\
Keck (1993)        &   3873   & 49  & 47 & 4 \\
Gran Telescopio Canarias (2007) & 3457 & 24 & 70 & 6 \\
\bottomrule
\end{tabular}
\end{table*}

\section{The Contemporary Observatory\label{sec:Contemporary}}

The development of cities from the nineteenth century began to cause problems in some nineteenth-century observatories, which not only began to withdraw from large cities to smaller settlements but also to seek locations far from urban areas. This withdrawal became increasingly necessary in the following decades, underlining a physical distance from the ordinary 
citizen that also led to an intellectual disconnection from astronomers until the press/media started to recover the connection amongst them in the 1960s.
Astronomical centers with spaces such as conference halls, classrooms or exhibitions, open to visitors, amateurs and experts, or even public observatories, lost prominence in favor of highly specialized scientific complexes almost exclusively for astronomy.

The search for the highest possible precision from the instruments resulted in optimization of the observation conditions, which led to the establishment of these institutions in areas not affected by anthropogenic (light) pollution and with more favorable atmospheric conditions. The Lick Observatory (1881) was the first continuously 
inhabited mountain observatory in the world \citep{smiley1938,misch1998}. 
The location of the center was fundamentally based on its astronomical advantages, without regard of the distance to any city or university 
\citep{neubauer1950}.
The architectural characteristics of the building are more like those of modern observatories than other contemporary observatories, which is why the Lick Observatory represents an intermediate, transitional case between the two epochs.

Almost simultaneously, the Nice Observatory (1881) emerged as a set of 
independent structures scattered over the top of a mountain. Each  
building was designed and located for a specific function, including
the first building devoted entirely to a telescope. In this way, 
it pioneered the contemporary observatory layout of a set of specialized and 
independent buildings \citep{holden1891,etienne2014,hunsch2012}.

The size of the new large reflectors, the differences in scale, the different needs intrinsic to the multiple uses or the orography of the mountainous terrain that makes it difficult to 
have wide level soils, are aspects that drive the segregation of the initially unified observatory building into a series of independent structures. 
This separation allows for a freedom of disposition and orientation in the territory that will favor the conditions of observation, minimizing interference (vibrations, smoke, light, heating) that would compromise the quality of the astronomical observation.

Although mirror-based telescopes had been installed in observatories for decades, their role had always been secondary to that of large refractors. However, their importance increased thanks to 
technical advances that enabled the manufacture of the required mirrors in parallel with the development of astrophysics. In addition, the dissociation of the observatory buildings allowed the adaptation of the constructions to the instruments they housed without external constraints. An example of this is the Mount Wilson Observatory (1904), initially harboring the large 60-inch 
reflector and, later, in the 100-inch Hooker telescope 
\citep{adams1954}.

Ignoring the heterogeneous large-scale layout of the modern observatory, the contemporary design for an individual instrument is generally a base architectural structure scaled to the instrument it contains, topped by a movable cover, generally a dome. The size of telescopes reached such magnitudes that it rivaled the very scale of the architecture that had to contain it. This meant increasing the size of the buildings that housed the 
telescopes, producing a large amount of contained space available for other uses. Similarly, these large instruments require major structural supports, to such an extent that spaces inside the 
supports themselves can accommodate some functions. 

Following the earlier Hooker telescope development, the Hale Telescope (1948) 
ended up consolidating different functions and hosting 
almost a thousand people during its opening ceremony. In this way, these large telescopes would incorporate or could incorporate enough secondary 
functions (offices, work canteens, meeting rooms, laboratories, warehouses, workshops, rest rooms, toilets ...) to be able to contain practically 
everything required for an entire observatory. Thus, the construction planned for each instrument would have sufficient autonomy to constitute an astronomical observatory by itself.

\begin{figure*}[t!]
\centering
\includegraphics[angle=0,width=160mm]{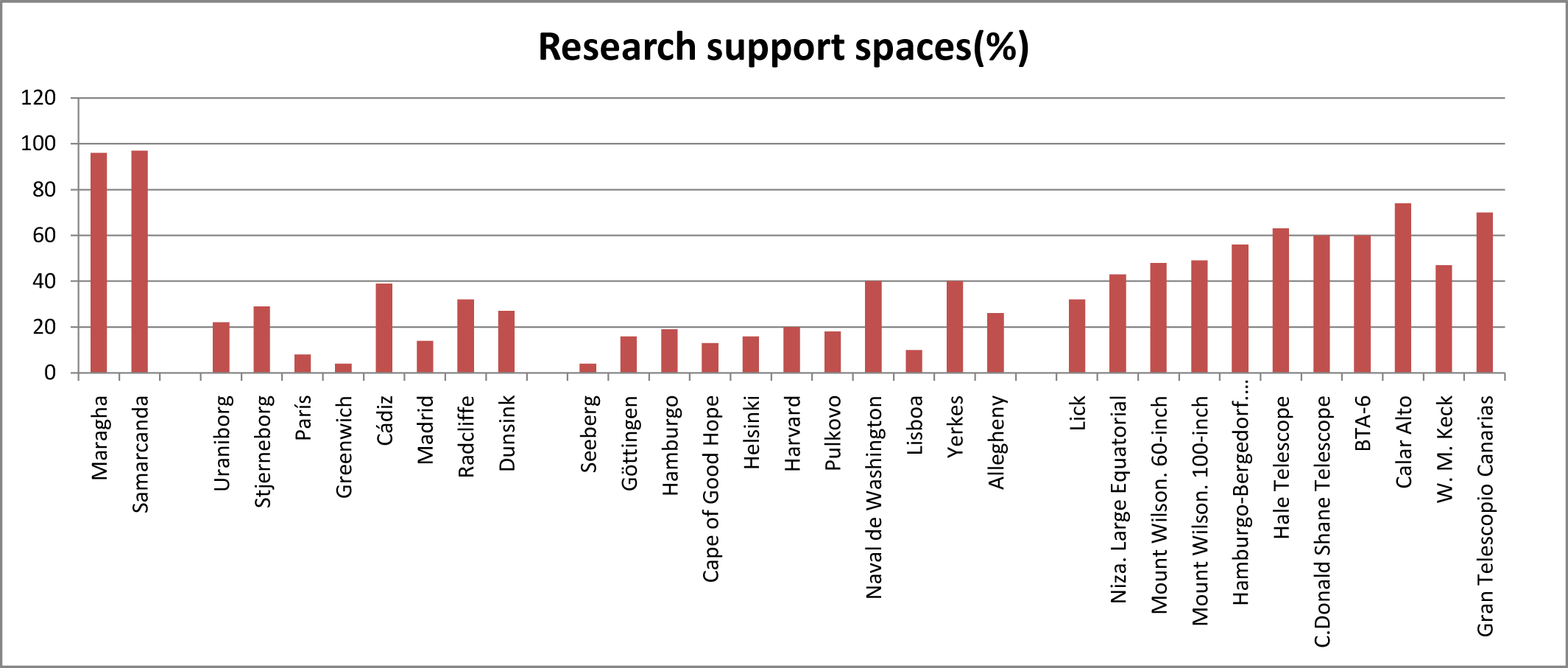}
\caption{Fractions of the surface areas allocated to research support spaces in the observatories in our sample.\label{fig:jkasfig2}}
\end{figure*}

From this point on, the most important observatories were constituted as sets of independent structures where, habitually, they would include buildings for some accessory uses 
(residences, visitor areas ...), other structures allocated to secondary instruments, and the main buildings housing the great telescopes together with the primary functions (meeting rooms, work rooms, laboratories, offices ...), which would constitute the essence of the observatory.

The constant growth of the mirrors caused mechanical-structural problems in the BTA-6 (1975) (with a 6-meter primary mirror) that forced to introduce some fundamental changes in its design, adopting an alt-azimuth instead of an equatorial mount. This variation complicated the tracking of celestial objects, requiring a computerized control system. These two revolutionary advances were consolidated and replicated in later observatories \citep{leverington2017}.
After the largest monolithic mirrors (8.2-meter diameter) were produced for the European Southern Observatory's Very Large Telescope (VLT, first light in 1998) 
\citep{schilling1998} and Subaru (8.2~m), and the Large Binocular Telescope (8.4~m) later on, the planners of the W.~M. Keck Observatory 
(1993/1996) decided to design the primary mirror as a combination of smaller mirrors to make ever-increasing mirror sizes possible. 
Correctly aligning segmented mirrors at that time represented a very hard problem. 
Because of this, for VLT a single-mirror approach was selected. Keck did indeed have control problems for years, vindicating the VLT decision.  However, the experience of fixing Keck's problems paved the way for the future larger segmented mirrors telescopes.

The alt-azimuth design allowed for a significant reduction of the structures, which could be light, and the size of the domes, which would also reduce their cost \citep{finn1985} compared to equatorially mounted telescopes. 
The race to the largest reflector paved the way to the Gran Telescopio Canarias (GTC, first light in 2007), with effective aperture of 10.4~m (composed of 36 segmented mirrors of 1.8~m diameter each) and inspired by the Keck. However, other projects are already underway that will exceed it in size but without significant changes in its configuration: the GMT (24.5m), the TMT (30m) or the ELT (39m).

Despite the changes experienced during this period, or even from the very origin of this institution, the importance of technological development in terms of computerization and communication systems must be highlighted as the causes of the greatest observatory revolution. Both the remote control and the remote access to the observations mark a decisive turning point regarding these complexes, since it seems to offer the possibility of abandoning all physical presence in the observatories. And although it is not necessary for an astronomer to look through a lens to access an observation, there are still a series of tasks derived from it that require a space in which scientists and engineers can work.

\section{Space-function Analysis\label{sec:Analysis}}

Beyond this summary of the development of the observatory up to the present, anticipating the future evolution needs some analysis. 
Since it is difficult to evaluate all the variables that affect the constitution of these astronomical centers, this study 
focuses on the configuration of its architecture in relation to its uses, considering the evolution of astronomy and astronomical observing as a function of time.
For this purpose, a classification is established for the different activities for which the different spaces of the buildings that house the instruments are intended. 
Observation spaces are considered to be those places occupied by active observation equipment, that 
is, they can be in operation if the environmental conditions make it possible. 
Research support spaces are those space that, though not directly participate in observations, provide functions necessary for astronomical work. Included in this category are work rooms, workshops, instrument warehouses, and so on.
Finally, all of the dependencies that are not essential for astronomical research, such as conference rooms, rest areas, visitor galleries or exhibitions, are described as service spaces.

Based on this classification, we selected a representative sample 
(Figure~\ref{fig:jkasfig1}) of astronomical centers that had the greatest impact and influence on other observatories of their time and afterwards. They are grouped according to the 
periods described in the previous sections and ordered chronologically and summarized in Table~\ref{tab:jkastable1}, which is based on the 
analysis of the plans and drafts of the different observatories together with information on how various spaces were used.

\begin{figure*}[t!]
\centering
\includegraphics[angle=0, width=160mm]{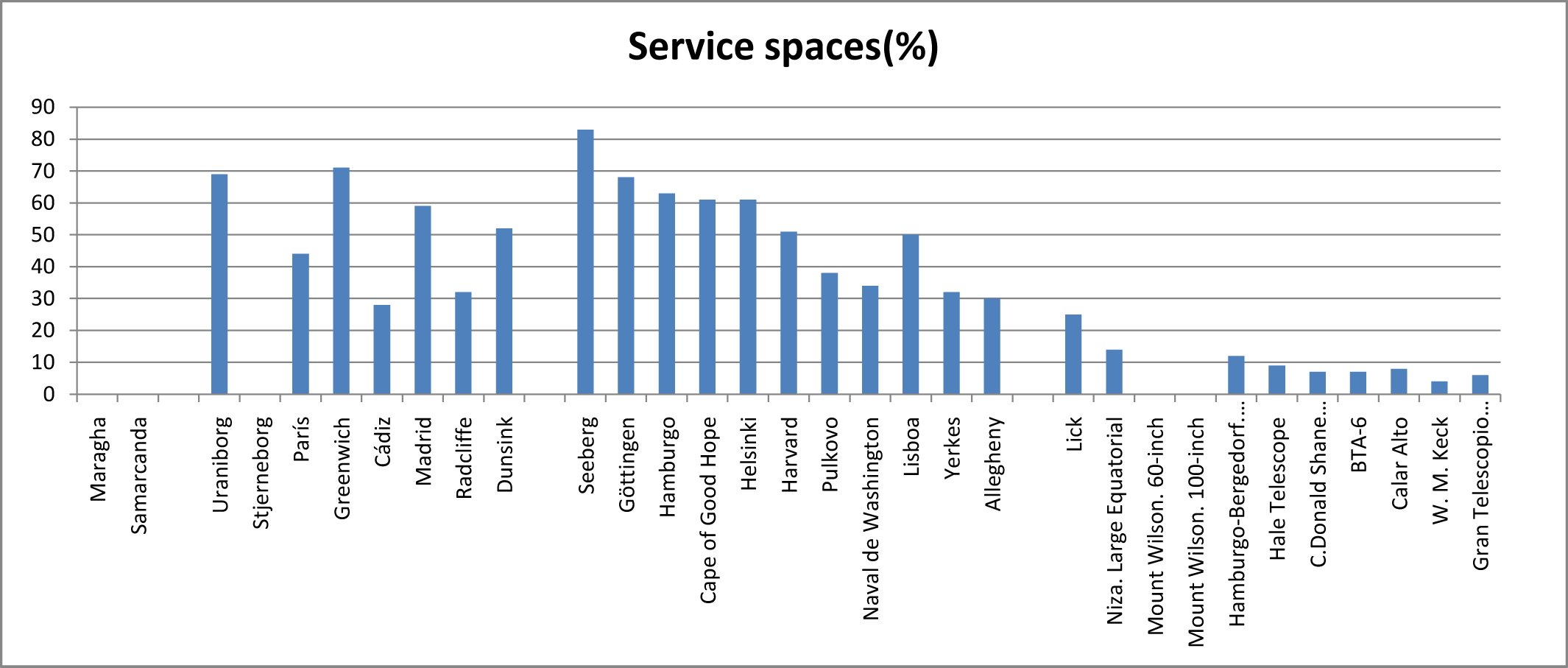}
\caption{Fractions of the surface areas dedicated to service spaces in the observatories in our sample.\label{fig:jkasfig3}}
\end{figure*}

Our work shows how the aspects defined in the different periods have their corresponding manifestation in this classification. 
This highlights the weak correlation between the Islamic observatory and the modern observatory. 
In addition, during the beginnings of the modern observatory (between the Uraniborg and Dunsink cases) there is a period of continuous changes 
in the distribution of uses, probably related to the different types of instrumentation and observing techniques used. 
However, since its consolidation (since Seeberg) a tendency has emerged 
regarding the development of its organization that extends to the current contemporary observatory.

On the one hand, Figure~\ref{fig:jkasfig2} shows a tendency toward an increase of the fraction occupied by research support spaces over time.
That means a higher proportion of area in which astronomers and technicians can actually work (offices, laboratories, workshops ...), 
which suggests how the complexity of the instruments attached to the telescopes has increased as time went on. It underscores the growing importance of these functions for achieving results in research.

On the other hand, as shown in Figure~\ref{fig:jkasfig3}, the fraction occupied by service spaces decreases with time. This can be due to a greater specialization of these buildings as they suppress some functions not strictly astronomical (rest areas, cafeterias, exhibition halls or visitors centers) which may be relegated to secondary structures. 

An important reason for the scattered or segregated pattern of modern observatories is the addition of new projects after the initial foundation. Good sites are rare, getting water, electricity and roads to that 
site may be the single most expensive part of the project. By comparison, adding further instruments to an established site is often a minor expense.  Also, in terms of project planning, existing observatories have a science-grade record of the local weather, eliminating the need for possibly several years of site monitoring to establish whether a new proposed site is good enough.

\section{The Forthcoming Observatory \label{sec:Forthcoming}}

Since its inception, the astronomical observatory has reflected a continuous architectural adaptation of the building to the needs of astronomy. 
Thus, aspects such as structural stability, thermal stability, orientation of the construction, the specific configuration of windows, 
the adaptation to the different astronomical instruments, the generation of service and comfort spaces for the users of the observatory, the incorporation of new functions derived from the advances of science, the change of scale according to the growth of the instruments, the adaptation to non-urban geographic environments or the conciliation with new technologies have been addressed.

Obviously the current observatories, of the beginning of the 21st century, are vastly different from their modern or Islamic-medieval ancestors. Nor are they the same as those of the mid-twentieth century, even though the influence of the Hale Telescope has spread through all the astronomical centers that succeeded it. In any case, this evolution is not taking place in an intentional or planned way, but the changes take place in an organic and almost improvised way. This, which is not a problem in itself, does denote a lack of deep understanding regarding observatories, understood as buildings in which astronomical research and human activity coexist and relate through architecture. All this means that occasionally some of the different functions of the observatory come into conflict with each other, indicating a certain lack of criticism and reflection. It is difficult to answer clearly and unequivocally where astronomical observatories are heading, since there does not seem to be a roadmap that points in a clear direction, due to the fact that technological advances are unpredictable.

\section{Conclusions \label{sec:Conc}}

Although the selection of astronomical observatories analyzed in this work could include a larger sample, it would not significantly change the result since the architecture and configuration of the buildings was frequently replicated with few variations. 

In any case, our investigation has found evidence for two opposite directions of evolution:
\begin{itemize}
\item continuation of the dissociation of astronomical instruments from other functions related to research;
\item return to heterogeneous uses of space in the same building such that the needs of its users can match the astronomical activity.
\end{itemize}

The first of these possibilities would involve a split of what is now known as an observatory. The result is a pair of buildings, one of which houses all the spaces and dependencies necessary for the staff to do their research (laboratories, offices, etc.), all the elements that could result in better comfort for workers (cafeteria, dining room, rest areas, etc.). In addition, this property should also house all kinds of technical rooms (control rooms, workshops ...) or accessories (facilities, warehouses ...) that would serve the astronomical function. The second building, separate and independent, would be completely unrelated to human activity (for example, the electricity supply).

The second possibility would involve unifying in a single building both the strictly intrinsic functions of astronomical research and the rest of the uses and services associated with the workers of the center or even visitors. For this, it would be necessary to carefully consider and correct the possible interference and incompatibilities between the different uses. This would result in a rather complex single construction.

When planning a new observatory in an area where other research centers are concentrated, it would be reasonable to choose a structure for the new telescope that would be separated from a place for workers and other users that could even be shared with other facilities and neighbors. However, opting for a unitary building would be adequate if the new observatory considers aspects beyond pure research (such as teaching or public outreach) or if the project were to be implemented in a remote and isolated place, saving resources and optimizing supply lines.

Overall it seems reasonable that the two opposite directions of evolution can coexist in the near future where they could represent the most appropriate solutions in different scenarios; see also \citet{castrotirado2019b}.


\acknowledgments

We have profited from valuable discussions with Ronan Cunniffe and Gemma Olmedo Paradas. We are grateful to Luis Alberto Rodr\'iguez and Antonio Cabrera-Lavers (Gran Telescopio Canarias), Cristina Rodr\'iguez (Observatorio de Sierra Nevada), Santos Pedraz (Calar Alto Observatory) and Tatyana Sokolova (Special Astrophysical Observatory) for providing the requested material and pictures of the astronomical observatories they are related to. We also acknowledge very useful comments from the two anonymous referees. M. A. C. T. has been supported by a Junta de Andaluc\'ia predoctoral grant under project TIC-2839. Finally, we also acknowledge financial support from the State Agency for Research of the Spanish MCIU through the ''Center of Excellence Severo Ochoa'' award for the Instituto de Astrof\'isica de Andaluc\'ia (SEV-2017-0709).


\end{document}